\documentclass{iaus}
\usepackage{graphicx}

\title[IAUS 277.~~Lyman Break Analogs] 
{Lyman Break Analogs: Constraints on the Formation of Extreme Starbursts at Low and High Redshift}

\author[Gon\c{c}alves, T. S.]   
{Thiago S. Gon\c{c}alves$^1$, Roderik Overzier$^2$, Antara Basu-Zych$^3$, \\D. Christopher Martin$^1$
}

\affiliation{$^1$California Institute of Technology \\ 1200 E. California Blvd. MC 278-17, Pasadena CA 91125, USA \\ email: {\tt tsg@astro.caltech.edu}\\[\affilskip]{$^2$Max-Planck-Institut for Astrophysics, D-85748 Garching, Germany\\[\affilskip]}
{$^3$NASA Goddard Space Flight Center, Laboratory for X-ray Astrophysics, Greenbelt, MD 20771, USA\\[\affilskip]}
}

\pubyear{2010}
\volume{277}  
\pagerange{1-4}
\jname{Tracing the Ancestry of Galaxies in the Land of Our Ancestors}
\editors{Carignan, C., Freeman, K.C. \& Combes, F., eds.}
\begin{document}

\maketitle

\begin{abstract}
Lyman Break Analogs (LBAs), characterized by high far-UV luminosities and surface brightnesses as detected by GALEX, are intensely star-forming galaxies in the low-redshift universe ($z\sim 0.2$), with star formation rates reaching up to 50 times that of the Milky Way. These objects present metallicities, morphologies and other physical properties similar to higher redshift Lyman Break Galaxies (LBGs), motivating the detailed study of LBAs as local laboratories of this high-redshift galaxy population. We present results from our recent integral-field spectroscopy survey of LBAs with Keck/OSIRIS, which shows that these galaxies have the same nebular gas kinematic properties as high-redshift LBGs. We argue that such kinematic studies alone are not an appropriate diagnostic to rule out merger events as the trigger for the observed starburst. Comparison between the kinematic analysis and morphological indices from HST imaging illustrates the difficulties of properly identifying (minor or major) merger events, with no clear correlation between the results using either of the two methods. Artificial redshifting of our data indicates that this problem becomes even worse at high redshift due to surface brightness dimming and resolution loss. Whether mergers could generate the observed kinematic properties is strongly dependent on gas fractions in these galaxies. We present preliminary results of a CARMA survey for LBAs and discuss the implications of the inferred molecular gas masses for formation models.
\end{abstract}

\firstsection 
\section{Introduction}

The paradigm for the formation of galaxies has changed dramatically over the course of the last decade. New integral field units (IFUs) have allowed the detailed study of star-forming galaxies at $z\sim 2$, when the universe was young and forming the bulk of its current stellar mass. These studies (Law et al. 2009, F\"orster-Schreiber et al. 2009) have revealed new details about the properties of the nebular gas in starburst galaxies, showing that these objects, unlike local star-forming spirals, present strong random dynamical components, with velocity dispersion reaching values well above 100 km s$^{-1}$.

However, the interpretation of these data is difficult, mainly because of the loss of surface brightness due to cosmic dimming and the loss of spatial resolution. Therefore, the importance of a comparison sample at low redshift, one with similar physical properties but less subject to observational biases, becomes evident.

In an attempt to find such a sample, Heckman et al. (2005) and Hoopes et al. (2007) have selected the galaxies in the low redshift universe ($z\sim 0.2$) with highest values of FUV luminosity and surface brightness ($L_{FUV} > 2\times 10^{10} L_\odot$, $I_{FUV} > 10^9 L_\odot$ kpc$^{-2}$), based on GALEX data and with thresholds determined as to mimic the properties of typical Lyman Break galaxies (LBGs) at $z\sim 2-3$. Even though these were the only selection criteria, subsequent studies have shown that these galaxies strongly resemble LBGs in most of their physical properties, including metallicity, radio, morphology, extinction and outflow properties (Hoopes et al. 2007, Basu-Zych et al. 2007, Overzier et al. 2009, Overzier et al. 2010, Overzier et al. 2011, Heckman et al. 2011). Therefore, we refer to this sample  as the Lyman Break analogs (LBAs).

In this work we describe our IFU observations of the emission line gas kinematics in LBAs, comparing our results to those found for high-redshift star-forming galaxies. These results, along with a detailed description of the observations, are presented in full in Basu-Zych et al. (2009) and Gon\c{c}alves et al. (2010).

\section{Results}

We have observed 19 LBAs with the OSIRIS instrument at Keck. The compact sizes and strong hydrogen line emission of these galaxies are ideal for this purpose, since we can achieve high signal-to-noise ratios and still fit within the small field of view of the instrument. By using laser guide star adaptive optics, we are able to resolve approximately 200 pc in each galaxy, which represents the diffraction limit of Keck at these redshifts. We have also artificially redshifted our data to $z\sim 2$, to allow for a direct comparison with the results from the aforementioned IFU studies at high redshift selected from the literature.

The first important result is that the gas kinematics in LBAs is similar to that found in the star-forming galaxies at high-redshift. We also find velocity dispersion values above 100 km s$^{-1}$ in some cases, which generate low $v/\sigma$ values, that is, the ratio between velocity shear and velocity dispersion. In Figure 1, we show these values as a function of stellar mass. Intrinsic values are slightly higher than typical ratios found at high redshift (black triangles), but once we redshift our data, we lose information at the outer radii, where the gas is most offset from the central velocity. The result is that $v/\sigma$ appears smaller (red and blue symbols), with values similar to those found in Law et al. (2009) and F\"orster-Schreiber et al. (2009) (hollow symbols). We also show that intrinsic values depend on stellar mass, with more massive galaxies presenting higher $v/\sigma$ ratios, indicating a stronger rotational dynamical component, as observed also at $z\sim 2-3$.

Furthermore, we have analyzed the asymmetry of the kinematic maps, as described by Shapiro et al. (2008) using VLT/SINFONI data. This kinemetry technique consists of decomposing the velocity and velocity maps into Fourier components in order to compare the purely symmetrical component with higher-order ones. Shapiro et al. (2008) have used this technique to differentiate between rotating disks (low asymmetry) and mergers (high asymmetry). We find that galaxies with higher stellar masses present lower asymmetry values, reinforcing the idea of more massive galaxies more closely resembling local spirals. In addition, we have shown that the loss of spatial resolution at high redshift makes galaxies appear smoother, rendering asymmetry values smaller; in this way, 50\% of our galaxies would be wrongly classified as relatively undistorted disks at higher redshift. This is shown in Figure 1, where we compare intrinsic asymmetry values with those determined for the artificially redshifted observations. Approximately half the galaxies fall inside the gray-shaded area, indicating galaxies that would be classified as ``mergers'' at low-z but as ``disks'' at high-z.

\begin{figure}[b]
\begin{center}
 \includegraphics[width=2.3in]{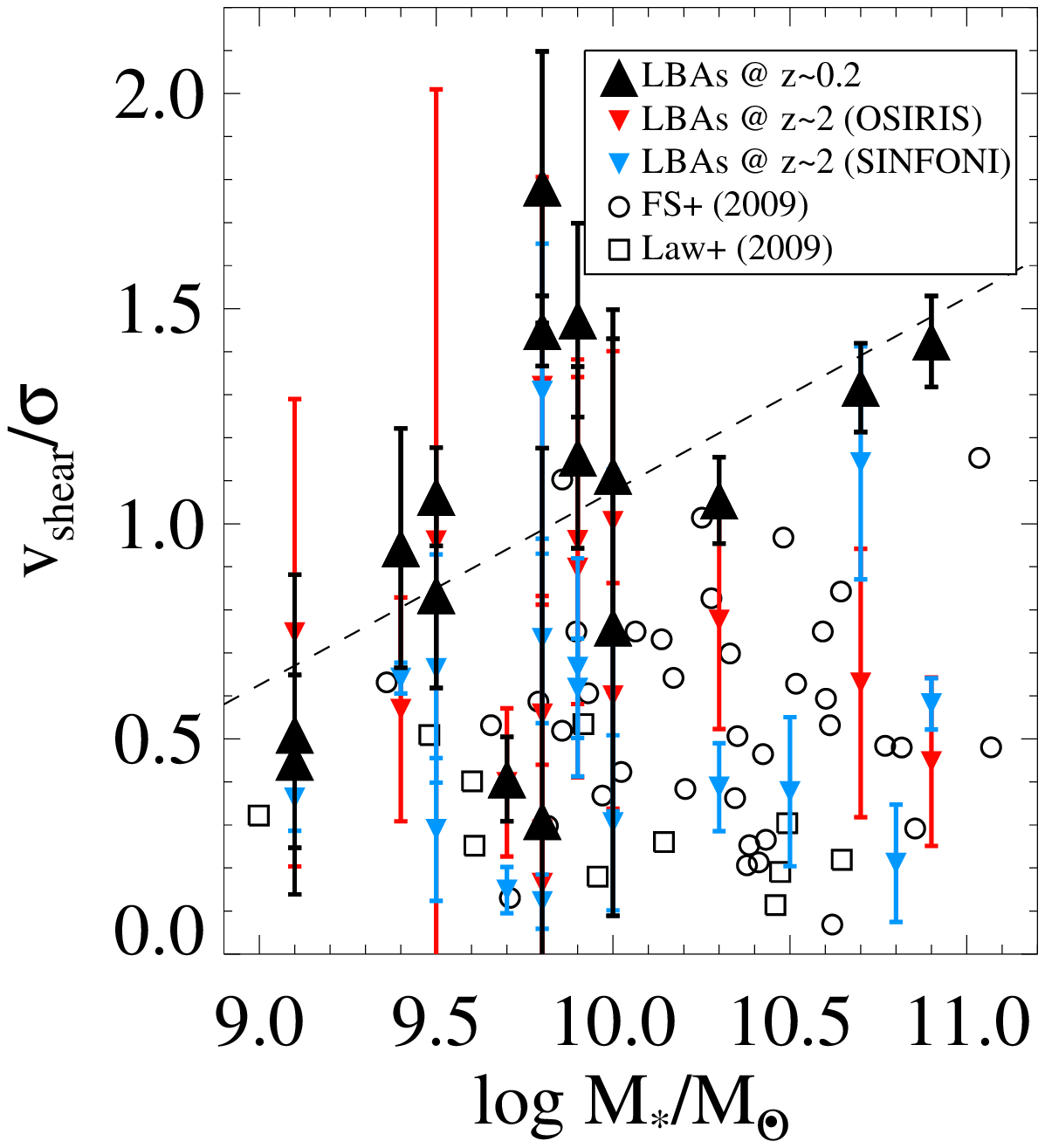} 
 \includegraphics[width=2.3in]{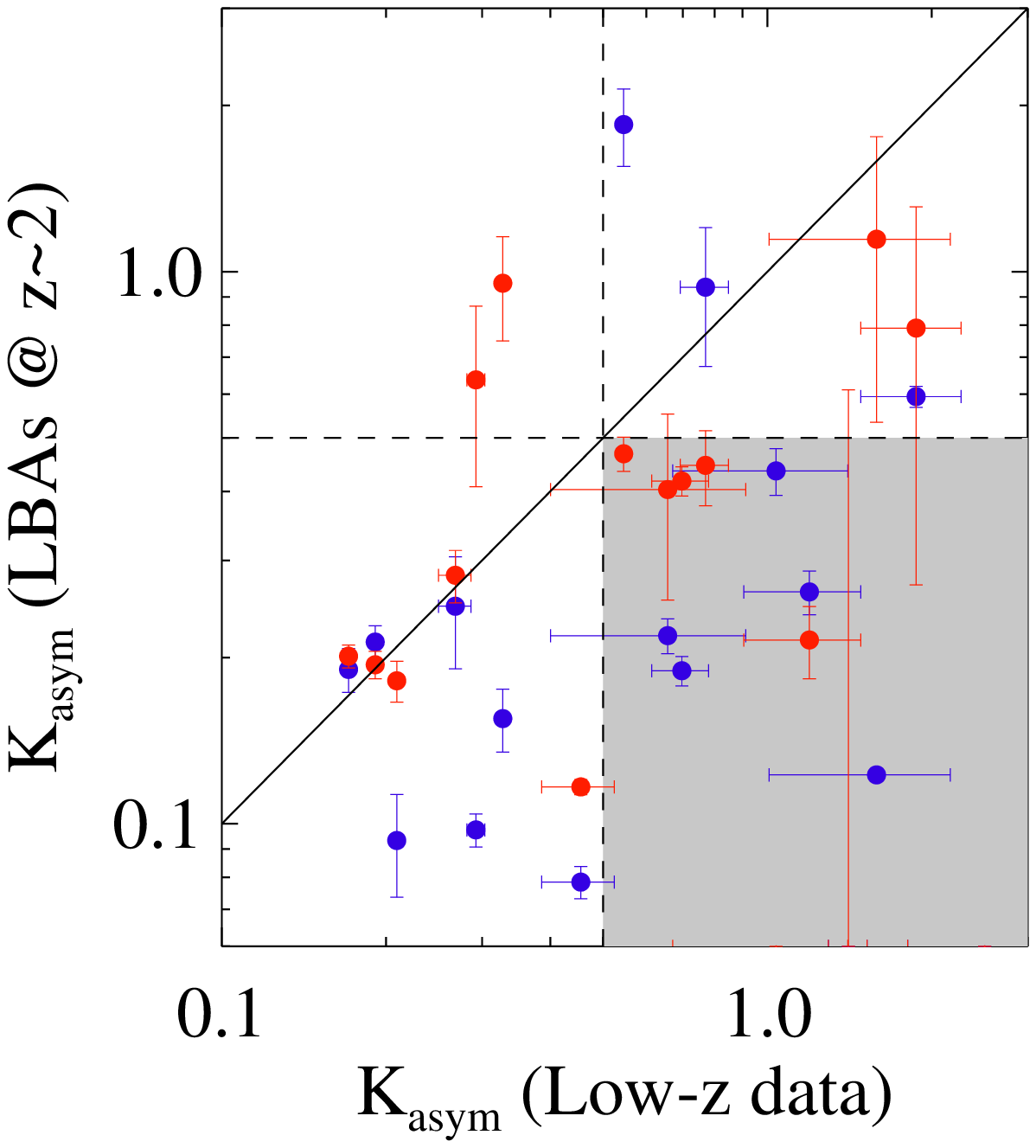} 
 \caption{{\it Left}: $v/\sigma$ ratios for starburst galaxies. Black triangles indicate intrinsic values for our LBA data, while downward blue and red triangles represent the artificially redshifted data. Hollow symbols indicate actual high-redshift data from Law et al. (2009) and F\"orster-Schreiber et al. (2009). The dashed line is a fit to the intrinsic LBA data, evidencing the stellar mass dependence. {\it Right}: Comparison of kinematic asymmetry indices between intrinsic ($x$-axis) and artificially redshifted data ($y$-axis). Dashed lines represent the threshold between mergers and disks as in Shapiro et al. (2008). The gray shaded area shows galaxies classified as ``mergers'' at low redshift but as ``disks'' at high $z$. Most points lie below the solid line, indicating that galaxies look less asymmetric at high redshift. Figures from Gon\c{c}alves et al. (2010).}
   \label{fig1}
\end{center}
\end{figure}

Particularly striking examples of these effects are shown in Gon\c{c}alves et al. (2010, see Figure 10 of that work), where we show a case having the expected kinematic profile of a rotating disk, while careful examination of the HST $r$-band morphology reveals clear signs of a recent merger event as evidenced by clumpy starburst, tidal tails, and an asymmetric outer isophote. Another example is shown in which evidence for a rotating disk only appears after redshifting the LBA to $z\sim 2$, while no such evidence exists in the full resolution, full sensitivity data. In addition to these technical problems, there are also problems related to the timescale of merger events. For example, Robertson \& Bullock (2008) have shown an example of a gas-rich merger remnant rapidly cooling and forming a rotating disk that would be difficult to identify as such through IFU observations.

\section{Molecular Gas}

The amount of cold molecular gas in the reservoir available for star formation is a crucial ingredient in understanding the starbursts in star forming galaxies, and in particular whether a merger could generate a rotating disk in such short timescales. With this in mind, we have initiated a survey searching for CO(1-0) emission in LBAs. We have thus far been awarded 92 hours of time with the CARMA array. We present preliminary results in the following paragraphs.

Out of three objects targeted, we have two detections and one upper limit, after approximately 4 hours of integration in each case. As further evidence for the analogy between our sample and high redshfit LBGs, we present in Figure 2 the gas mass fraction of galaxies as a function of stellar mass. We note that the expected gas fractions in LBAs from the Schmidt-Kennicutt relation are very similar to expected values for LBGs. Moreover, these expectations are confirmed by actual CARMA measurements: gas fractions vary between $\sim 50$\% at intermediate stellar masses to $\sim 10-20$\% at the high-mass end.

\begin{figure}[t]
\begin{center}
 \includegraphics[width=2.4in]{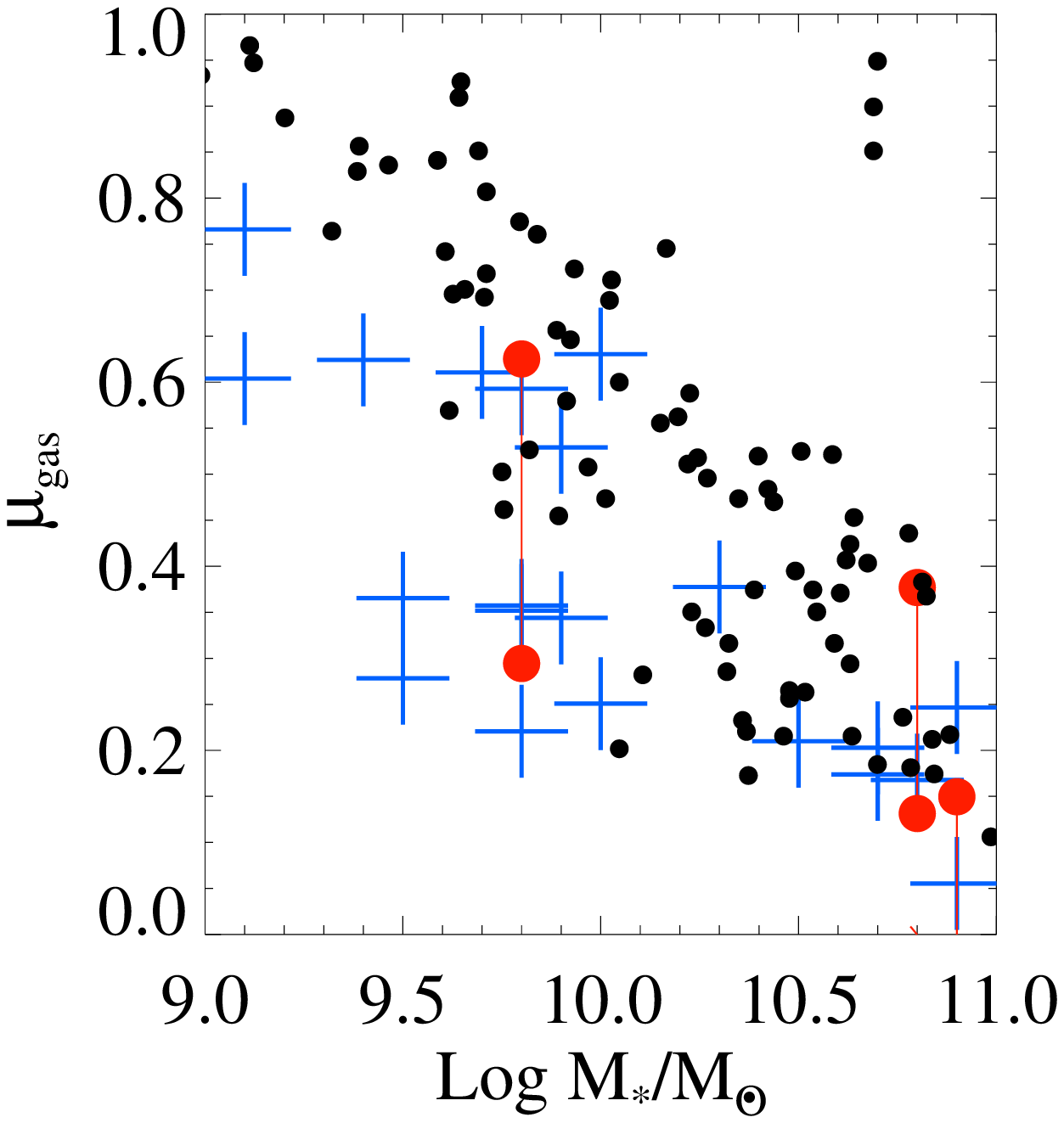} 
 \includegraphics[width=2.4in]{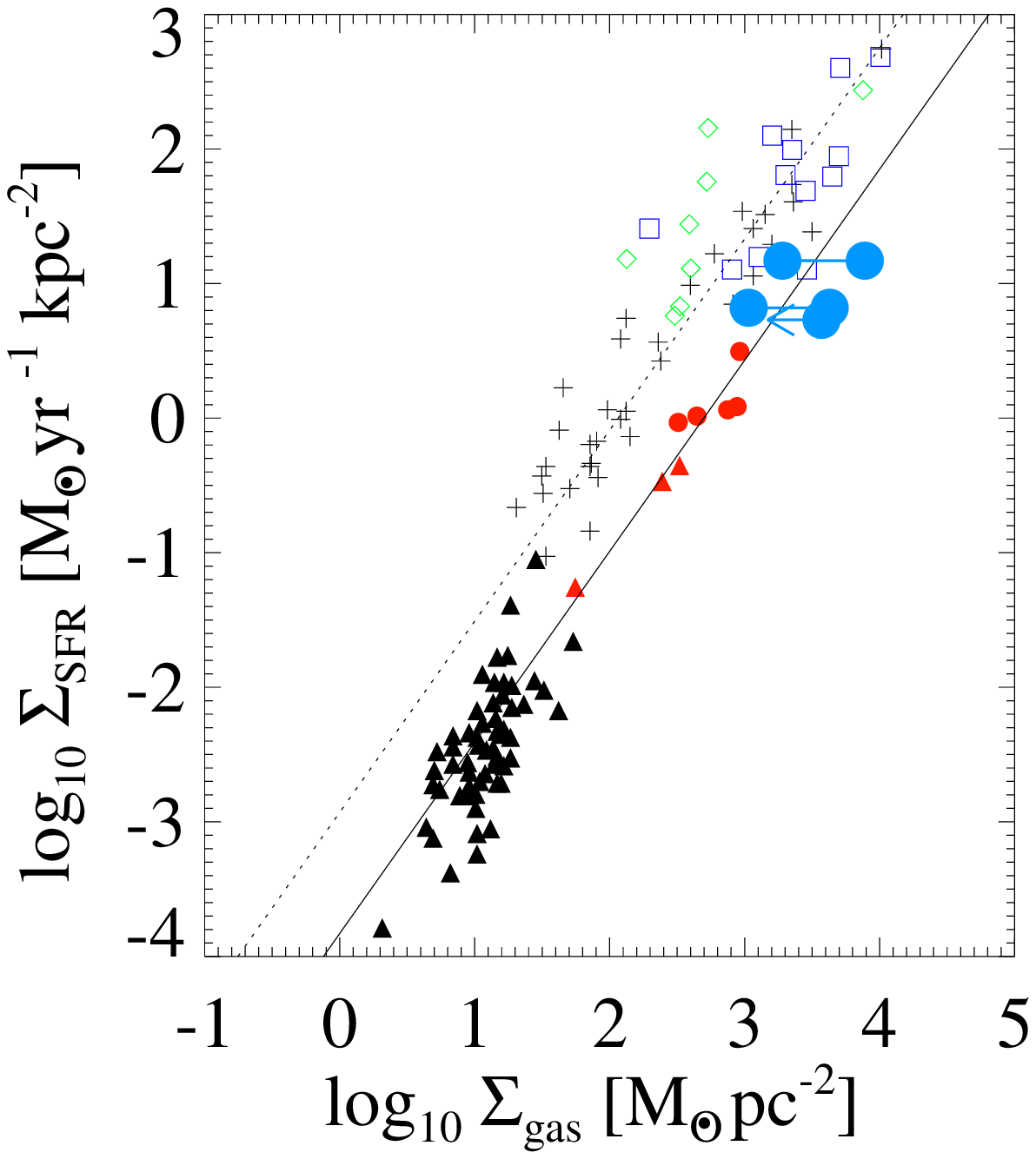} 
 \caption{{\it Left}: Gas fractions for LBGs (black dots; Erb et al. 2006) and LBAs (blue crosses) as calculated from the Schmidt-Kennicutt relation, as a function of stellar mass. The red symbols represent actual CARMA measurements; we conservatively use a range of CO-to-gas conversion of $0.9 \leq \alpha \leq 4.2$, hence the range in $\mu_{gas}$ values. {\it Right}: The Schmidt-Kennicutt relation for different galaxy samples. LBAs (blue circles) follow the relation for local spirals. For more details on all samples, see Daddi et al. (2010).}
   \label{fig2}
\end{center}
\end{figure}

Another interesting finding is that LBAs appear to follow the local Schmidt-Kennicutt relation, similar to what is found for BzK galaxies at $z\sim 2$ by Daddi et al. (2010), only at higher surface densities. This reinforces the idea of a bimodal Schmidt-Kennicutt relation, since our sample is distinct from more infrared bright galaxies such as ULIRGs and SMGs, which present a higher star formation surface density for a given surface density of gas (see Daddi et al. 2010 and references therein).

With strong CO(1-0) line emission (with fluxes up to $\sim 5$ Jy km s$^{-1}$), LBAs will be ideal candidates for further studies with ALMA once the instrument comes online. Not only will detections be possible within minutes, but we will be able to measure the surface density of gas at a detail level only achievable for local galaxies so far. We will thus be able to measure the spatially resolved Schmidt-Kennicutt relation at scales below 100 pc, providing invaluable insight into the interplay between the gas reservoir and star formation in starburst galaxies.

\end{document}